\documentclass[pra,twocolumn,amsmath,amssymb]{revtex4}
\usepackage{graphicx}
\usepackage{bm,fancybox}
\usepackage{amsmath,amssymb}
\usepackage{color}
\usepackage{ascmac}

\begin{document}
\title{Acausal measurement-based quantum computing}  
\author{Tomoyuki Morimae}
\email{morimae@gunma-u.ac.jp}
\affiliation{ASRLD Unit, Gunma University,
1-5-1 Tenjin-cho Kiryu-shi Gunma-ken, 376-0052, Japan}
\date{\today}
            
\begin{abstract}
In the measurement-based quantum computing, there is a natural ``causal cone"
among qubits of the resource state, since the measurement angle
on a qubit has to depend on previous measurement results
in order to correct the effect of byproduct operators.
If we respect the no-signaling principle, byproduct operators cannot be avoided.
In this paper, we study the possibility
of acausal measurement-based quantum computing
by using the process matrix framework 
[O. Oreshkov, F. Costa, and C. Brukner, Nature Communications {\bf3}, 1092 (2012)].
We construct a resource process matrix for 
acausal measurement-based quantum computing
restricting local operations to projective measurements.
The resource process matrix
is an analog of
the resource state of the standard causal measurement-based quantum
computing.
We find that if we restrict local operations to projective measurements
the resource process matrix is (up to a normalization factor
and trivial ancilla qubits) 
equivalent to the decorated graph state 
created from the graph state of the
corresponding causal measurement-based quantum computing.
We also show that it is possible to consider a causal game whose causal inequality
is violated by acausal MBQC.
\end{abstract}

\pacs{03.67.-a}
\maketitle  

\section{Introduction}
In Ref.~\cite{Oreshkov}, 
Oreshkov, Costa, and Brukner
proposed a novel framework, which is called the process matrix (PM) framework, to study
general physics on multipartite systems where locally quantum physics is assumed but
globally no restriction, such as the no-signaling and the causality,
is set (see also Refs.~\cite{Brukner1,Brukner2,SWolf,SWolf2}). 
They showed that this framework can describe general theory beyond
the standard quantum physics, including a ``mixture" of different
time causal orders.
Interestingly, they explicitly constructed an example of the PM system whose induced
correlation violates a ``causal inequality" that is satisfied
by all space-like and time-like correlations~\cite{Oreshkov}.

In this paper, we study measurement-based quantum computing (MBQC)~\cite{MBQC} in the
PM framework.
MBQC is a new model of quantum computing proposed by Raussendorf and Briegel.
In this model, universal quantum computation can be done
with only local measurements on each qubit of a certain quantum many-body
state, which is called a ``resource state".
While the computational power of MBQC is equivalent to the traditional circuit
model of quantum computation, MBQC provides novel view points
to deepen our understanding of quantum computing. In fact, 
plenty of new results have been obtained by using MBQC,
such as
relations of quantum computing to condensed matter physics~\cite{Verstraete,Gross_QCTN,
MiyakeAKLT,Cai,Miyake_edge,Miyake2dAKLT,Wei2dAKLT,Cai_magnet,fMBQC,
upload,stringnet,Bravyi,Nest,Nest2},
the fault-tolerant topological MBQC~\cite{Raussendorf_topo,
Sean,FujiiTokunaga,Ying,Ying2,FMspin},
roles of quantum properties (such as entanglement, correlation, and purity)
in MBQC~\cite{SE1,SE2,Gross_ent,Bremner,Morimae_ent_fidelity,Morimae_mixed},
and secure cloud quantum computing~\cite{BFK,FK,Barz,Vedran_coherent,
AKLTblind,topoblind,CVblind,topoveri,MABQC,Sueki,composable,
composableMA,distillation,Lorenzo,Joe_intern,Barz2,honesty}. 

One of the most peculiar things in MBQC is that there is a
natural ``causal cone" among qubits of the resource state~\cite{Rau1,Rau2,Elham1,Elham2,Elham3}.
The measurement angle of a qubit has to be determined by the previous measurement results,
since we have to correct the effect of the byproduct operators,
which cannot be avoided (if we respect the
no-signaling~\cite{byproduct}).
Given the PM framework, it is natural to ask
``can we describe acausal MBQC in the PM framework?" 
Here, acausal MBQC means that the measurement angle of each qubit
does not depend on measurement results of other qubits, but
we can perform correct quantum computing.
In the PM framework, a density matrix is generalized
to a PM. Therefore, the above question is restated as
follows: ``can we find a resource PM (which corresponds to
a resource state of the causal MBQC) for acausal MBQC?"

The purpose of the present paper is to answer the question.
We explicitly construct a resource PM for acausal MBQC
restricting local operations to projective measurements.
In this acausal MBQC, the measurement angle of each qubit
can be independent from measurement results of other qubits.
Interestingly, if we restrict local operations to
projective measurements, the resource PM is (up to a normalization factor
and trivial ancilla qubits) equivalent to the decorated graph state 
of the corresponding causal MBQC.
(Here, a decorated graph state is a graph state whose graph is created
from the original graph by adding an extra vertex to each vertex of the original graph.)
We also consider a causal game whose causal inequality
is violated by acausal MBQC.

\section{PM framework}
Let us quickly review the PM framework.
Let us consider bipartite system, Alice and Bob.
(Generalizations to multipartite systems are straightforward.)
Alice is in her laboratory, which is isolated from the outer world.
In her laboratory, physics is governed by the quantum theory.
This means that an Alice's measurement corresponding to the
result $a$ is represented by a completely-positive (CP) trace-non-increasing map 
$
{\mathcal M}_a^A:{\mathcal L}({\mathcal H}^{A_1})\to{\mathcal L}({\mathcal H^{A_2}}),
$
where ${\mathcal H}^{A_1}$ and ${\mathcal H}^{A_2}$ 
are Alice's input and output Hilbert spaces, respectively,
${\mathcal L}({\mathcal H})$ is the space of operators
over a Hilbert space ${\mathcal H}$,
and $\sum_a{\mathcal M}_a^A$ is a CP and trace-preserving (CPTP) map.
In a similar way, Bob is in his isolated laboratory, and
inside of the laboratory the quantum theory is correct.
His measurement corresponding to the result $b$ is represented
by a CP trace-non-increasing map 
$
{\mathcal M}_b^B:
{\mathcal L}({\mathcal H}^{B_1})\to{\mathcal L}({\mathcal H^{B_2}}),
$
where again $\sum_b{\mathcal M}_b^B$ is a CPTP map.
In this way, Alice's and Bob's local systems are explained in the
quantum theory.
However, no restriction is set on the physics of the outer world
where their laboratories are embedded.
In particular, the no-signaling and the causality are not assumed between
the two laboratories.
It was shown~\cite{Oreshkov}
that the probability $P({\mathcal M}_a^A,{\mathcal M}_b^B)$ that Alice's measurement result is $a$
and Bob's measurement result is $b$ is given by
\begin{eqnarray*}
P({\mathcal M}_a^A,{\mathcal M}_b^B)=\mbox{Tr}\Big[W^{A_1,A_2,B_1,B_2}
(M_a^{A_1,A_2}\otimes M_b^{B_1,B_2})\Big],
\end{eqnarray*}
where $W^{A_1,A_2,B_1,B_2}\in{\mathcal L}
({\mathcal H}^{A_1}
\otimes{\mathcal H}^{A_2}
\otimes{\mathcal H}^{B_1}
\otimes{\mathcal H}^{B_2})$, and
\begin{eqnarray*}
M_a^{A_1,A_2}&\equiv&\Big[(I\otimes {\mathcal M}_a^A)|ME\rangle\langle ME|\Big]^T\\
&=&\sum_{i,j=1}^d|i\rangle\langle j|\otimes{\mathcal M}_a^A(|j\rangle\langle i|)
\in{\mathcal L}({\mathcal H}^{A_1}\otimes{\mathcal H}^{A_2})
\end{eqnarray*}
is the positive semi-definite operator 
obtained by the
Choi-Jamiolkowsky (CJ) isomorphism.
Here, $d$ is the dimension of ${\mathcal H}^{A_1}$, $T$ is the matrix transposition,
and 
$
|ME\rangle\equiv\sum_{j=1}^d|j\rangle\otimes|j\rangle\in
{\mathcal H}^{A_1}\otimes{\mathcal H}^{A_1}
$
is the (non-normalized) maximally-entangled state.
The operator $M_b^{B_1,B_2}\in{\mathcal L}({\mathcal H}^{B_1}\otimes{\mathcal H}^{B_2})$ 
is defined in a similar way.
A map ${\mathcal M}^A\equiv\sum_a{\mathcal M}_a^A$ is CPTP if and only if
its CJ operator $M^{A_1,A_2}$ satisfies $M^{A_1,A_2}\ge0$
and $\mbox{Tr}_{A_2}M^{A_1,A_2}=I$.
If $W^{A_1,A_2,B_1,B_2}$
satisfies
\begin{eqnarray}
W^{A_1,A_2,B_1,B_2}\ge0
\label{W1}
\end{eqnarray}
and
\begin{eqnarray}
\mbox{Tr}\Big[W^{A_1,A_2,B_1,B_2}
(M^{A_1,A_2}\otimes M^{B_1,B_2})\Big]=1
\label{W2}
\end{eqnarray}
for all $M^{A_1,A_2}$ and $M^{B_1,B_2}$
such that
$M^{A_1,A_2}\ge0$,
$M^{B_1,B_2}\ge0$,
$\mbox{Tr}_{A_2}M^{A_1,A_2}=I$,
and $\mbox{Tr}_{B_2}M^{B_1,B_2}=I$,
we call $W^{A_1,A_2,B_1,B_2}$ the process matrix (PM)~\cite{Oreshkov}.
A PM is, in some sense, 
a generalization of a density matrix in quantum theory.
(If operation on $A_2$ and $B_2$ are identity, then the PM becomes density matrix.)

\section{MBQC}
\begin{figure}[htbp]
\begin{center}
\includegraphics[width=0.4\textwidth]{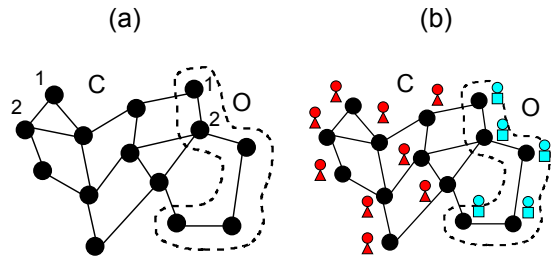}
\end{center}
\caption{(a) The resource state $\sigma$. For example,
it is the graph state $|G\rangle$ on the graph $G$. 
(b) The distributed MBQC.
Red people are Alice$_j$ ($j=1,...,N$)
and blue people are Bob$_j$ ($j=1,...,$).
}
\label{sigma}
\end{figure}

Before describing our result, we also review the basics of MBQC.
Let $\sigma$ be an $(N+n)$-qubit resource state of MBQC.
We divide $\sigma$ into two subsystems $C$ and $O$ (Fig.~\ref{sigma} (a)).
The subsystem $C$ consists of $N$ qubits,
and the subsystem $O$ consists of $n$ qubits. 
Qubits in $C$ are measured in order to
implement the ``C"omputation. The ``O"utput of the computation is encoded
on qubits in $O$, and therefore we measure qubits in $O$ 
to readout the output of the computation.
Measurements on $C$ are adaptive:
we first measure the first qubit of $C$ in a certain orthonormal
basis $\{|\phi_1^0\rangle,|\phi_1^1\rangle\}$.
Let $m_1\in\{0,1\}$ be the result of the measurement.
We next define an orthogonal basis, $\{|\phi_2^0(m_1)\rangle,|\phi_2^1(m_1)\rangle\}$,
which depends on $m_1$,
and measure the second qubit of $C$ in this basis.
If the measurement result is $m_2\in\{0,1\}$, we measure the third qubit
in the orthonormal
basis $\{|\phi_3^0(m_1,m_2)\rangle,|\phi_3^1(m_1,m_2)\rangle\}$,
and so on.
In this way, we adaptively measure all qubits in $C$.
After measuring all qubits in $C$,
we finally measure each qubit of $O$ in the computational basis 
$\{|0\rangle,|1\rangle\}$
in order to readout the computation result.
Depending on the measurement results on $C$, some operators (usually Pauli operators)
are acted upon $O$. Such operators are called byproduct operators.
Because of the effect of the byproduct operators,
the result on $O$ must be postprocessed.

The canonical example of the resource state is the graph state~\cite{MBQC}.
Let us consider a graph $G=(V,E)$ of $N$ vertices. 
The graph state $|G\rangle$ corresponding to the graph $G$
is defined by
$
|G\rangle\equiv\Big(\bigotimes_{e\in E}CZ_e\Big)|+\rangle^{\otimes N},
$
where $|+\rangle\equiv\frac{1}{\sqrt{2}}(|0\rangle+|1\rangle)$, 
and $CZ_e\equiv|0\rangle\langle0|\otimes I+|1\rangle\langle1|\otimes Z$
is the Controlled-Z gate between two vertices of the edge $e$.

\begin{figure}[htbp]
\begin{center}
\includegraphics[width=0.4\textwidth]{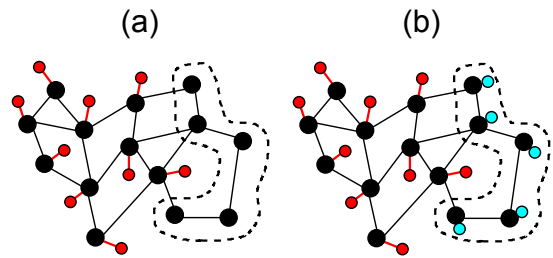}
\end{center}
\caption{(a) The decorated graph $G'$ created from the graph $G$ of Fig.~\ref{sigma} (a).
(b) Each blue circle is the completely-mixed state $\frac{I}{2}$. 
The entire state is now $|G'\rangle\langle G'|\otimes(\frac{I}{2})^{\otimes n}$.
} 
\label{sigma2}
\end{figure}



\section{Resource PM for acausal MBQC}
Now we show the main result.
Our acausal MBQC is performed in the distributed way (Fig.~\ref{sigma} (b))
by $N$ girls,
Alice$_j$ $(j=1,...,N)$, and $n$ boys,
Bob$_j$ $(j=1,...,n)$. 
They share a certain (possibly super-quantum) resource system consists
of $N+n$ particles. The system is divided into two subsystems
$C$ and $O$, which consist of $N$ and $n$ particles, respectively.
Alice$_j$ possesses $j$th particle of $C$,
and Bob$_j$ possesses $j$th particle of $O$ (Fig.~\ref{sigma} (b)).

In the causal MBQC, Alice$_j$ has to know the measurement results
of Alice$_k$ $(k=1,...,j-1)$ in order to determine her measurement angle.
However, in the present acausal MBQC, we assume that
Alice$_j$ measures her system in the fixed
orthonormal basis $\{|\phi_j^0\rangle,|\phi_j^1\rangle\}$
irrespective of the measurement results of Alice$_k$ $(k\neq j)$
and Bob$_j$ ($j=1,...,n$), 
where $|\phi_j^m\rangle\equiv\frac{1}{\sqrt{2}}(|0\rangle+(-1)^me^{i\phi_j}|1\rangle)$.
In the causal MBQC, we can no longer perform correct quantum computing
if Alice$_j$'s measurement is fixed in this way.
However, we will see later that in the acausal MBQC, we can perform correct
quantum computing
in spite that Alice$_j$'s measurement is fixed.

After the Alice$_j$'s measurement, Alice$_j$ sets the system to $|m_j\rangle$,
where $m_j\in\{0,1\}$ is Alice$_j$'s measurement result.
The CJ operator of such a measurement process is given by
\begin{eqnarray*}
\sum_{k=1}^d\sum_{l=1}^d
|k\rangle\langle l|\otimes |m_j\rangle\langle\phi_j^{m_j}|l\rangle
\langle k|\phi_j^{m_j}\rangle\langle m_j|
=\ovalbox{$\phi_j^{m_j}$}\otimes\ovalbox{$m_j$}.
\end{eqnarray*}
Here, we have used the convenient notation
$\ovalbox{$x$}\equiv|x\rangle\langle x|$~\cite{Vedran}.
Bob$_j$ measures his system in the computational basis $\{|0\rangle,|1\rangle\}$,
and sets the system to $|z_j\rangle$ after Bob$_j$'s measurement,
where $z_j\in\{0,1\}$ is Bob$_j$'s measurement result.
The CJ operator corresponding to Bob$_j$'s measurement is
thus $\ovalbox{$z_j$}\otimes\ovalbox{$z_j$}$.

Let us consider the decorated graph $G'$ (Fig.~\ref{sigma2} (a)) of the graph $G$,
which is created by adding an extra vertex to each vertex of $C$ in Fig.~\ref{sigma} (a).
We denote the graph state on the decorated graph $G'$ by $|G'\rangle$.
We also add $n$ completely-mixed states $(\frac{I}{2})^{\otimes n}$
to $|G'\rangle$ as is shown in Fig.~\ref{sigma2} (b).
Now we have the $(2N+2n)$-qubit state $\ovalbox{$G'$}\otimes(\frac{I}{2})^{\otimes n}$.
We claim that if we restrict local operations to projective measurements
the (unnormalized) $(2N+2n)$-qubit
state 
\begin{eqnarray*}
W&\equiv&2^{N+n}\ovalbox{$G'$}\otimes \Big(\frac{I}{2}\Big)^{\otimes n}\\
&=&2^{N+n}\Big(\bigotimes_{i=1}^N V_i\Big)
\Big(\ovalbox{$G$}\otimes\ovalbox{+}^{\otimes N}\Big)
\Big(\bigotimes_{j=1}^N V_j\Big)\otimes \Big(\frac{I}{2}\Big)^{\otimes n}
\end{eqnarray*}
is a resource PM for acausal MBQC corresponding
to the causal MBQC on $|G\rangle$. 
Here
$V_j$ is the Controlled-Z gate, 
$
V\equiv\ovalbox{0}\otimes I+\ovalbox{1}\otimes Z, 
$
between $j$th black qubit in $C$ and $j$th red qubit $|+\rangle$ (indicated by
red lines) of Fig.~\ref{sigma2} (a).
Note that $W$ satisfies Eq.~(\ref{W1}) since $W$ is nothing but an
unnormalized quantum state.
We will see later that Eq.~(\ref{W2}) is also satisfied for measurements
used in MBQC.

The probability of obtaining the measurement results 
$(m_1,...,m_N,z_1,...,z_n)\in\{0,1\}^{N+n}$
by Alice$_j$ ($j=1,...,N$) and Bob$_j$ ($j=1,...,n$)
in the acausal MBQC
is then given by
\begin{widetext}
\begin{eqnarray*}
&&P(\phi_1^{m_1},...,\phi_N^{m_N},z_1,...,z_n)\\
&=&
\mbox{Tr}\Big[W\times 
\Big(\bigotimes_{s=1}^N\ovalbox{$\phi_s^{m_s}$}\otimes\ovalbox{$m_s$}\Big)
\otimes
\Big(\bigotimes_{t=1}^n\ovalbox{$z_t$}\otimes\ovalbox{$z_t$}\Big)
\Big]\\
&=&
2^{N+n}\mbox{Tr}\Big[
\Big(\bigotimes_{i=1}^N V_i\Big)
\Big(\ovalbox{$G$}\otimes\ovalbox{+}^{\otimes N}\Big)
\Big(\bigotimes_{j=1}^N V_j\Big)
\otimes
\frac{I^{\otimes n}}
{2^n}
\times 
\Big(\bigotimes_{s=1}^N\ovalbox{$\phi_s^{m_s}$}\otimes\ovalbox{$m_s$}\Big)
\otimes
\Big(\bigotimes_{t=1}^n\ovalbox{$z_t$}\otimes\ovalbox{$z_t$}\Big)
\Big]\\
&=&
\sum_{(m_1',...,m_N')\in\{0,1\}^N}
\sum_{(m_1'',...,m_N'')\in\{0,1\}^N}
\mbox{Tr}\Big[
\Big(\bigotimes_{i=1}^NZ_i^{m_i'}\Big)
\ovalbox{$G$}
\Big(\bigotimes_{j=1}^NZ_j^{m_j''}\Big)
\otimes
|m_1',...,m_N'\rangle
\langle m_1'',...,m_N''|\\
&&\times
\Big(\bigotimes_{s=1}^N\ovalbox{$\phi_s^{m_s}$}\otimes\ovalbox{$m_s$}\Big)
\otimes
\Big(\bigotimes_{t=1}^n\ovalbox{$z_t$}\Big)
\Big]\\
&=&
\mbox{Tr}\Big[
\Big(\bigotimes_{i=1}^NZ_i^{m_i}\Big)
\ovalbox{$G$}
\Big(\bigotimes_{j=1}^NZ_j^{m_j}\Big)
\times
\Big(\bigotimes_{s=1}^N\ovalbox{$\phi_s^{m_s}$}\Big)
\otimes
\Big(\bigotimes_{t=1}^n\ovalbox{$z_t$}\Big)
\Big]\\
&=&
\mbox{Tr}\Big[\ovalbox{$G$}\times 
\Big(\bigotimes_{s=1}^N\ovalbox{$\phi_s^0$}\Big)
\otimes
\Big(\bigotimes_{t=1}^n\ovalbox{$z_t$}\Big)
\Big].
\end{eqnarray*}
\end{widetext}
In this way, irrespective of the measurement results $(m_1,...,m_N)$ on $C$,
we can always obtain the result of the causal MBQC
in the positive branch, i.e., all measurement results are correct 
$m_j=0$ ($j=0,...,N$).

Equation~(\ref{W2}) is satisfied for measurements used in MBQC, 
since
\begin{eqnarray*}
&&\sum_{m\in\{0,1\}^N}
\sum_{z\in\{0,1\}^n}
P(\phi_1^{m_1},...,\phi_N^{m_N},z_1,...,z_n)\\
&=&
\sum_{m\in\{0,1\}^N}
\sum_{z\in\{0,1\}^n}
\mbox{Tr}\Big[\ovalbox{$G$}\times 
\Big(\bigotimes_{s=1}^N\ovalbox{$\phi_s^0$}\Big)
\otimes
\Big(\bigotimes_{t=1}^n\ovalbox{$z_t$}\Big)
\Big]\\
&=&
2^N\mbox{Tr}\Big[\ovalbox{$G$}\times 
\Big(\bigotimes_{s=1}^N\ovalbox{$\phi_s^0$}\Big)
\otimes I^{\otimes n}
\Big]
=1,
\end{eqnarray*}
where $m\equiv(m_1,...,m_N)$,
$z\equiv(z_1,...,z_n)$, and
we have used the fact that every branch of measurement histories
occurs with the same probability in MBQC~\cite{MBQC}.

In Ref.~\cite{byproduct} it was shown that the byproduct operators
cannot be avoided if we respect the no-signaling principle.
This is because, if we can avoid byproducts, a person who possesses $C$ can create
any state in $O$, and if another far separated person possesses $O$, then the
first person can transmit information to the second person by encoding his message
in the created state.
Therefore, the acausal MBQC considered in this paper
should be in a class of signaling theory in the PM framework.
In fact, in the acausal MBQC considerd in this paper we can always create the correct
output quantum state in $O$ without any byproduct operators, 
since we can choose the correct branch.
If girls encode a message in the output quantum state, 
boys can always learn the message by measuring their system.
This means that the no-signaling from girls to boys is violated.

\section{MBQC Causal game}
We can consider the following causal game whose causal inequality is violated
by the acausal MBQC.
Let us again consider the distributed MBQC in Fig.~\ref{sigma} (b).
Let $P_0$ be the probability of obtaining the all zero result $0...0$
for Bob$_j$ ($j=1,...,n$).
In the causal MBQC, $P_0\le\frac{1}{2}(1+\frac{1}{2^n})$,
because if all girls are causally past to all boys, and
all girls are correctly ordered, then girls can steer boys' systems into the
state $|0\rangle^{\otimes n}$ up to the byproduct operators.
If girls send the measurement result to boys, boys can correct the byproduct operators,
and can obtain $|0\rangle^{\otimes n}$.
On the other hand, if all boys are causally past to all girls, and
all girls are ordered correctly, then boys' systems are the $n$-qubit
completely-mixed state, and therefore the probability of obtaining the all zero result
$|0\rangle^{\otimes n}$ is $\frac{1}{2^n}$.
As we have seen in the previous section, however, if we consider acausal MBQC,
$P_0=1$, since all girls and boys can always perform correct MBQC.

\section{Conclusion and Discussion}
In this paper, we have considered acausal MBQC
in the PM framework.
Assuming local operations are projective measurements,
we have constructed a resource PM for acausal
MBQC, and show that it is (up to a normalization factor
and trivial ancilla qubits) equivalent to the decorated graph state 
created from
the graph state of the corresponding causal MBQC.


Our result also suggests that acausal MBQC can be simulated
on a causal MBQC with postselection (postselecting red qubits in Fig.~\ref{sigma2} (b)).
Since the simulation of the postselected MBQC is possible for
a small size MBQC, we might be able to experimentally simulate
acausal MBQC on a small resource state.
(Since the success probability exponentially decreases, larger systems
would be hard to simulate.)
Such an approach will be connected to recently developed
important topics, namely, quantum simulations of phenomena beyond
quantum physics~\cite{sim}.
It would be interesting to further explore relations to the result.

In this paper we have considered only qubit graph state MBQC
with projective measurements.
It would be a future research subject to generalize
the present result to more general MBQC including local POVM
measurements.

We finally mention that quantum computing without definite causal order
was also studied in the circuit model with ``quantum switch"~\cite{Qswitch}.
They provided an example of quantum computing which cannot be implemented by
inserting a single use of black box in a quantum circuit with fixed time order.
Such quantum computing offer some advantages, such as black box discrimination 
problems~\cite{QS1}
and reducing an unknown black box query complexity~\cite{QS2}.
Since circuit model with projective measurements are equivalent to MBQC,
it would be interesting future study to consider
relations between the present result and quantum switch.

\acknowledgements
The author is supported by the Tenure Track System MEXT Japan
and the KAKENHI 26730003 by JSPS.

\end{document}